\documentstyle[12pt]{article}

\begin{document}
\title{Cosmic String Wakes in Scalar-Tensor Gravities}
\author{S. R. M. Masalskiene\thanks{sandra@mat.unb.br} \hspace{0.1mm} \mbox{and} M. E. X. Guimar\~aes\thanks{emilia@mat.unb.br} \\
\mbox{\small{Universidade de Bras\'{\i}lia, Departamento de Matem\'atica}} \\
\mbox{\small{Campus Universit\'ario, CEP. 70910-900}}
\mbox{\small{Bras\'{\i}lia, DF, Brazil}}}
\date{}
\maketitle
\begin{abstract}
The formation and evolution of cosmic string
wakes in the
framework of a scalar-tensor gravity are investigated in this work.
We consider a simple model in which cold dark matter flows past an
ordinary string and we treat this motion in the Zel'dovich approximation.
We  make a comparison between our results and previous results obtained
in the context of General Relativity. We propose a mechanism in which the 
contribution of the scalar field to the evolution of the wakes may lead to 
a cosmological observation.
\end{abstract}

\section{Introduction}

Topological defects can be formed as a result of one or
several thermal phase transitions which occurred in the Early Universe
\cite{kibble}. In particular, much attention has been given to cosmic strings'
models since they are possible sources for the density perturbations which
seeded galaxy formation.
A relevant mechanism to understand the structure formation by cosmic
strings
involves long strings moving with relativistic speed in the normal plane,
giving rise to velocity perturbations in their wake. If the string is moving with normal velocity $v_s$
through matter, a velocity perturbation
$u=8\pi G\mu v_s \gamma$
(where $\mu$ is the linear mass density of the string and
$\gamma =(1 -v_s^2)^{-1/2}$) towards the plane behind the string results
\cite{vil}.
The study of the effects of a cosmic string passing through matter
is of great importance to understand the current organization of matter
in the Universe and in this context, many authors have already
considered this problem in General
Relativity [3-8].

On the other hand, it is to notice that all implications of
structure formation by cosmic strings have been done in the framework of General Relativity. Nevertheless, it is generally believed  that gravity may not be
described by Einstein's action at sufficiently
high energy scales where
gravity becomes scalar-tensorial in nature. Indeed,
most attempts to unify
gravity with the other interactions predict the
existence of one (or many)
scalar(s) field(s) with gravitational-strength coupling. If gravity is essentially scalar-tensorial, there will be direct implications for cosmology and experimental
tests of the gravitational phenomena. In particular, all of them will be affected
by the variation of the gravitational ``constant" $\tilde{G}_{0}$.

Topological defects have been already considered in the context of a
scalar-tensorial gravity. In particular, the authors in ref.
\cite{gund, br} have studied the solutions for domain walls and  cosmic strings in
the simplest scalar-tensorial gravity - the Brans-Dicke theory;
in ref. \cite{gregory} the authors have studied the cosmic string in the
dilaton gravity, and finally, in ref. \cite{guima}, the author has
considered a local cosmic string model in more general scalar-tensor
theories in which the conformal factor is an arbitrary function of the scalar field. More
recently, superconducting strings have been considered in Brans-Dicke \cite{sen} and
in more general scalar-tensor theories \cite{mexg}.

In this work, we consider the formation and evolution of cosmic strings
wakes in  a scalar-tensor  gravity. Namely, we are interested in studying the
implications of a scalar-tensor coupling for the formation and evolution of
wakes by a cosmic string with metric described in  \cite{guima}.
For this purpose, we consider a simple model in which non-baryonic
cold dark matter composed by non-relativistic, collisionless particles
propagate around a scalar-tensor cosmic string. We anticipate that our
main result is to show that the mechanism of formation and evolution of
wakes by an (ordinary) cosmic string in the framework of a scalar-tensor
theory presents very similar structure to the
same mechanism by a wiggly
cosmic string in General Relativity.

This work is outlined as follows. In section 2, we briefly review the
properties of the metric of a scalar-tensor string which we are going
to deal with  throughout this paper
and we derive the linearized geodesic equations associated to it.
In section 3, we consider the propagation of non-relativistic,
collisionless
particles in the metric described in section 2 and we study the formation and
evolution of wakes in this model. Whenever
convenient, we reduce our results to the particular case of Brans-Dicke theory
and we compare these results with those already obtained in the framework
of General Relativity. Finally, in section 4, we end with some discussions
and conclusions.

\section{Cosmic String Solution in Scalar-Tensor Theories}

We  start by considering a class of scalar-tensor theories developed by
Bergman \cite{bergmann}, Wagoner \cite{wagoner} and Nordtverdt \cite{nordtverdt}
in which the  scalar sector of the gravitational interaction is massless.
For technical purpose, it is better to work in the so-called Einstein (conformal) frame in which the kinematic terms
of tensor and scalar fields do not mix. Then, a cosmic string solution arises from the action: 
\begin{eqnarray}
\label{3}
S & = & \frac{1}{16 \pi G_*} \int \mbox{d}^4 x \sqrt{-g}
\left[ R - 2 g^{\mu \nu} \partial_{\mu} \phi
\partial_{\nu} \phi \right] + \nonumber \\
& &  \int d^4 x \sqrt{-g} \left[ \frac{1}{2}D_{\mu}\Phi D^{\mu}\Phi^* -
\frac{1}{4}F_{\mu\nu}F^{\mu\nu} - V(\mid \Phi \mid) \right]  ,
\end{eqnarray}
where $g_{\mu \nu}$ is a pure rank-2 metric tensor, $R$ is the curvature scalar associated to it and $G_*$ is some ``bare" gravitational coupling constant. The second term in the r.h.s. of eq. (1) 
is the matter action representing an Abelian-Higgs model where a charged scalar Higgs field $\Phi$ minimally couples to the $U(1)$ gauge field $A_{\mu}$ and 
$V(\mid \Phi \mid)$ is the Higgs potential \cite{guima}. 

Action (1) is obtained from the original action appearing in the refs. 
[15-17] by a conformal transformation (see, for instance, \cite{dam})
\begin{equation}
\label{4}
\tilde{g}_{\mu \nu} = A^{2}(\phi) g_{\mu \nu} \;\; ,
\end{equation}
where $\tilde{g}_{\mu\nu}$ is the physical metric which contains both 
scalar and tensor degrees of freedom,  
and by a redefinition of the quantities
\[
G_*A^{2}(\phi) = \frac{1}{\tilde{\Phi}} \;\; ,
\]
where $\tilde{\Phi}$ is the original scalar field, and
\[
\alpha(\phi) \equiv \frac{\partial \ln A}{\partial \phi} = \frac{1}{[2 \omega(\tilde{\Phi}) + 3]^\frac{1}{2}}\;\; ,
\]
which can be interpreted as the (field-dependent) coupling strenght between matter and the scalar field. We choose to leave $A^2(\phi)$ as an arbitrary function of the scalar field. 

The metric 
of a self-gravitating string can be found in the weak-field approximation.  
Expanding the tensor $ g_{\mu \nu} = \eta_{\mu \nu} + h_{\mu \nu}$ and 
scalar $ \phi = \phi_{0} + \phi_{(1)}$ fields, with $\mid h_{\mu \nu} \mid  \ll 1$ and $\left|\frac{\phi_{(1)}}{\phi_0}\right| \ll 1$, it was, then, found \cite{guima}:
\begin{equation}
\label{9}
\mbox{ds}^2 = \left[1 + 8 G_0 \mu \alpha^2(\phi_0) \ln \frac{\rho}{\rho_0} \right]\left[ \mbox{dt}^2 - \mbox{dz}^2-
\mbox{d}\rho^2 - (1- 8 G_0 \mu) \left( \frac{\rho}{\rho_0} \right)^{2} \mbox{d}\theta^2 \right]
\end{equation}
and
\[
\phi_{(1)} = 4 G_{0} \mu  \alpha(\phi_{0}) \ln \rho/\rho_0 ,
\]
in a cylindrical coordinate system such that $\rho \geq 0$ and 
$0 \leq \theta < 2\pi$. $G_0$ is defined as $G_0 \equiv G_* A^2(\phi_{0})$. Notice that this is {\em not} the effective Newtonian constant $\tilde{G}_0 = G_*A^2(\phi_0)[
1+\alpha^2]_{\phi_0}$ as defined in \cite{dam}. 
$\rho_{0}$ is a distance beyond which all matter fields drop away. Conveniently, it has the same order of magnitude of the string radius.
It is interesting to note that the metric of a scalar-tensor
string in the weak-field approximation (3) is fully
described in terms of  one dimensional coupling strength
$(G_{0})$ and one post-Newtonian parameter $(\alpha (\phi_0))$.
One new feature of a scalar-tensor string is that it exerts gravitational force on test particles, contrary to its General Relativity  partner. From (3), we can easily see that this gravitational force on a test particle of mass $m$ is given by
\begin{equation}
\label{10}
f= - 4 m G_{0} \mu \alpha^2(\phi_0) \frac{1}{\rho}
\end{equation}
and it is always attractive.

\subsection*{Propagation of Massive Particles and Light}

Let us define a new coordinate system by means of the transformation $
x  = \rho \cos[(1- 8 G_{0} \mu)^{\frac{1}{2}}
\theta + 4 \pi G_0 \mu]$ and $
y =  \rho \sin[(1- 8 G_{0} \mu)^{\frac{1}{2}}
\theta + 4 \pi G_0 \mu]$ in such a way that the missing wedge is placed on the 
positive side of the $x$-axis. 
Then, metric (3) assumes a simple form:
\begin{equation}
\label{12}
\mbox{ds}^2 = (1 + h_{00})[ \mbox{dt}^2 - \mbox{dx}^2 - \mbox{dy}^2 - \mbox{dz}^2]
\end{equation}
where $ h_{00} = 8 G_{0} \mu \alpha^2(\phi_0) \ln[(x^2 + y^2)^{\frac{1}{2}}]$.
Metric (5) is conformally Minkowskian and has a missing wedge of angular width $\Delta = 8 \pi G_0 \mu$.  

Now we are in a position to study the propagation of massive
and massless particles in metric (5). Since this metric
is conformal to Minkowski minus a wedge, any massless particles (such as photons) will be deflected by an angle equal to $8 \pi G_0 \mu$. 
From the observational point of view, it would be impossible to
distinguish a scalar-tensor string from its General Relativity
partner just by considering effects based on deflection of light (i.e., double image effect, for instance). 
On the other hand, trajectories of massive particles will be
affected by the scalar-tensor coupling (which generates the gravitational force (4)) as well as by the conical geometry. If the string is moving with normal velocity $v_s$ through matter, a velocity perturbation 
\begin{equation}
\label{14}
u = 8 \pi G_0 \mu v_s \gamma +
\frac{4 \pi G_0 \mu \alpha^2 (\phi_0)}{v_{s} \gamma},
\end{equation}
where $\gamma = (1 - v_{s}^{2})^{- \frac{1}{2}}$, towards the plane behind the string results \cite{guima}. 
The first term is equivalent to the relative velocity of
particles flowing past a string in General Relativity and is due to the conical geometry. The second term appears due to the
scalar-tensor coupling of  the gravitational interaction.

Let us make an estimative of the order of magnitude of the corrected term induced by the scalar field in expression (\ref{14}). It is very illustrative to  consider a particular form for the arbitrary function $A(\phi)$, corresponding to the Brans-Dicke theory, $A(\phi) = e^{\alpha \phi}$ , with $\alpha^{2} = \frac{1}{2 \omega + 3}, (\omega = cte)$.
In this case, we have that 
$ G_* A^{2}(\phi_{0})= G_{0}=
\left( \frac{2 \omega + 3}{2 \omega + 4} \right) G_{eff}$ \cite{bd} 
where $G_{eff}$ is the Newtonian constant. Therefore, metric (\ref{9}) reduces to
\[
ds^2 = \left[ 1 + \frac{8\mu G_{0}}{2\omega + 3}\ln\frac{\rho}{\rho_0} \right] [dt^2 - dz^2 - d\rho^2 - (1 - 8\mu G_0) d\theta^2 ] ,
\]
in agreement with the result previously obtained by  Barros and Romero \cite{br}.

In the Brans-Dicke case, the expression for the relative velocities of particles flowing past a string reduces to:
\begin{equation}
\label{15}
u = 8\pi \left(\frac{2\omega + 3}{2\omega +4} \right)G_{eff}\mu v_s \gamma + \frac{4\pi G_{eff}\mu}{(2\omega +4) v_s \gamma} .
\end{equation}
Using the values for $\omega$ such that $\omega > 2500$
(consistent with solar system experiments made by Very Long Baseline 
Interferometry (VLBI) \cite{eub}) and
$<v_s> \sim 0.15$ (consistent with strings simulations \cite{allen}),
we conclude that the first term is more than 230 times larger than the
second one in expression (7).

\section{Formation and Evolution of Wakes in Scalar-Tensor Gravities}

\subsection{The Formation of Wakes: Cold Dark Matter Flowing Past the String}

Matter through which a long string moves, acquires a boost (\ref{14})
in the direction of the surface swept out by the string. Matter moves
toward this surface by gravitational attraction, and a wake is formed
behind the string.
The aim of this section is to study the implication of a scalar-tensor coupling for the formation and evolution of a wake behind a string which generates the metric (\ref{12}). For this purpose, we will mimic this  situation with a simple model in which cold dark matter composed by non-relativistic collisionless particles moves past a long scalar-tensor string. In our approach, particles propagate in the plane orthogonal to the string in a region $ \rho \equiv \sqrt{x^{2} + y^2} \gg R_0$, where
$R_0$ is the size of a region beyond which the string's
small-structures do not affect the motion.

The geodesics associated to metric (5) 
\begin{eqnarray*}
2 \ddot{x} & = & -(1 - \dot{x}^2 - \dot{y}^2) \partial_{x}h_{00} \nonumber \\
2 \ddot{y} & = & -(1 - \dot{x}^2 - \dot{y}^2) \partial_{y}h_{00},
\end{eqnarray*}
where $( \cdot )$ refers to derivative with respect to the coordinate $t$,  
can be integrated over the unperturbed
trajectories $x= X_0 + v_s t$ and $y = y_0$ . To linear order in $G_{0} \mu$, we have
\begin{eqnarray}
\label{16}
& \dot{x} & = v_s - 2 G_0 \mu \alpha^2(\phi_{0})
\left( \frac{1- v_{s}^{2}}{v_{s}} \right) \ln \left[
\frac{x^2 + y_{0}^{2}}{X_{0}^{2} + y_{0}^{2}} \right ]
\nonumber \\
& \dot{y}& = - 4 G_{0} \mu \alpha^{2} (\phi_0)
\left(  \frac{1 - v_{s}^{2}}{v_s} \right) \left[
\arctan \frac{x}{y_{0}} - \arctan  \frac{X_{0}}{y_{0}} \right]
\end{eqnarray}
where the particle has started its motion at
$x= X_0$ and $y=y_0$ with initial velocity $(\dot{x} = v_ s)$
 at $t=t_0$. $X_0$ is a long distance cut-off of order of the interstring separation.

Integrating (\ref{16}) again and writing $y$ as function of $x$, we have:
\begin{equation}
\label{17}
y = y_0 - 4 G_0 \mu \alpha^2(\phi_0) \left( \frac{1 - v_{s}^{2}}{v_{s}^{2}} \right) \left\{ x \left[ \arctan \frac{x}{y_0} - \arctan \frac{X_0}{y_0} \right] + \frac{y_0}{2} \ln \left[ \frac{x^2 + y_{0}^{2}}{X_{0}^{2} + Y_{0}^{2}} \right] \right\}
\end{equation}
With the orbit given by (\ref{17}), we can infer how particles
accrete onto the wakes. This is better seen in polar coordinates $(\rho, \theta)$, such that $\rho = \sqrt{x^2 + y^2}$ and $\theta$ is the angle measured from the $x$-axis.
If we assume that particles moving toward the string come from both sides from impact parameters $\delta R$ and $\delta R^{'}$ and that the
number of particles crossing the element $\delta \rho$ into the
foward cone per unit time per unit length of string is
$\nu v_{s} \delta R$, with $\nu$ as the initial number density
of particles, then the number density of particles entering the cone at element $\delta \rho$ is
$n_e(\rho, \theta) = \nu \frac{\rho}{R} \frac{\delta R}{\delta \rho}$
and the number of particles leaving the cone is
$ n_l(\rho, \theta) = \nu \frac{\rho}{R^{'}}
\frac{\delta R^{'}}{\delta \rho}$ .
In both cases, we have used the conservation of angular
momentum in order to calculate the velocity normal to the surface of the cone $R v_{s}/ \rho$ \cite{vacha}.

If we re-write orbit (\ref{17}) in terms of the polar coordinates and differentiate with respect to $y_0$, we find:
\begin{equation}
\label{18}
\frac{\delta R}{\delta \rho} = \frac{R}{\rho}
\left [ 1 + 4 G_0 \mu \alpha^2(\phi_0)
\frac{\rho \cos \theta - X_0}{X_{0}^{2} + R^2} \right].
\end{equation}
Therefore, we can now compute the number density of particles in the volume element $\delta \rho$:
\begin{equation}
\label{19}
n \equiv n_e + n_ l = \nu \left[ 2 + 4 G_0 \mu
\alpha^2(\phi_0) X_0 (x- X_0) \left( \frac{1}{X^{2}_{0} + R^2} + \frac{1}{X^{2}_{0} + R^{'2}} \right) \right],
\end{equation}
which holds just inside the wake.

Let us now compute the density fluctuation inside and outside the wake. We have, then, respectively:
\begin{equation}
\label{20}
\frac{\delta n}{\nu} \equiv \frac{n - \nu}{\nu} \approx 1 + 8 G_0 \mu \alpha^2(\phi_0) \left( \frac{x - X_0}{X_0} \right),
\end{equation}
and
\begin{equation}
\label{21}
\frac{ \delta n_e}{\nu} \approx 4 G_0 \mu \alpha^2(\phi_0)
\left( \frac{x - X_0}{X_0} \right).
\end{equation}
In both cases, we restricted ourselves to $R, R^{'} \ll X_0$.

We can now compare our results with those already known in the litterature. In General Relativity, the orbit of a test particle is deviated from its unperturbed trajectory because of the conical geometry and , in this case, the two opposite streams of matter in the wake overlap within the
wedge with opening $8\pi G \mu$ and the inside matter density is
doubled. In a scalar-tensor gravity, this scenario is changed and some new effects occur. Namely, the particle's original trajectory is also
perturbed by the presence of the scalar field and a new term induced by the gravitational force (\ref{10}) appears. Outside the wake, since $x < X_0$, $\delta n_e$ given by (\ref{21}) is negative which means that this region is underdense. An important result of this analysis is that the structure
of the formed wake in scalar-tensor gravity is very similar to the one in  the case of a wiggly string in General Relativity \cite{vacha}. We
will explore this similarity in more detail in the section 4.

\subsection{The Evolution of Wakes: The Zel'dovich Approximation}

Let us now make a quantitative description of accretion onto wakes
using the Zel'dovich approximation, which consists in considering the Newtonian accretion problem in an expanding Universe using the method of linear perturbations. Let us consider that a wake is formed by the scalar-tensor string at $t_i > t_{eq}$. The physical trajectory of a dark-particle can be written as
\begin{equation}
\label{22}
h(\vec{x}, t) = a(t) [\vec{x} + \psi(\vec{x}, t)] \;\; ,
\end{equation}
where $\vec{x}$ is the unperturbed comoving position of the particle and $\psi(\vec{x}, t)$ is the comoving displacement developed as a consequence of the gravitational attraction induced by the wake on the particle. If we assume that the wake is perpendicular to the x-axis, then the only non-vanishing component of $\psi$ is $\psi_x$. Therefore, the equation of motion for a dark particle in the Newtonian limit is
\begin{equation}
\label{23}
\ddot{h}= - \nabla_h \Phi \;\; ,
\end{equation}
where the Newtonian potential satisfies the Poisson's equation:
\begin{equation}
\label{24}
\nabla^2_h \Phi = 4 \pi G_ 0 \rho \; .
\end{equation}
In equation (\ref{24}), $\rho(t)$ is the dark matter density in a
cold-dark matter Universe. Thus, the linearized equation for $\psi_x$ becomes
\begin{equation}
\label{25}
\ddot{\psi} + 2 \frac{\dot{a}}{a} \dot{\psi} + 3\frac{\ddot{a}}{a} \psi = 0 \; .
\end{equation}
For simplicity, we consider hereafter that the Universe is flat.
Therefore $a(t) \propto t^{\frac{2}{3}}$ in the matter-dominated era such that $t> t_{eq}$, and equation ({\ref{25})
becomes:
\begin{equation}
\label{26}
\ddot{\psi} + \frac{4}{3t} \dot{\psi} - \frac{2}{3 t^2} \psi = 0 \;\; ,
\end{equation}
with appropriate initial conditions: $\psi(t_i) =0 $ and 
$ \dot{\psi}(t_i) = -u_i $. Equation ({\ref{26}}) is the Euler equation whose solution can be
easily found:
\begin{equation}
\label{28}
\psi (x,t) = \frac{3}{5}\left[ \frac{u_i t_{i}^{2}}{t} - u_i t_i \left(\frac{t}{t_i} \right)^{\frac{2}{3}} \right] \;\; .
\end{equation}
The comoving coordinate $x(t)$ can be calculated using the fact that $\dot{h} =0$ in the ``turn around''. That is, eventually,
the dark particle stops expanding with the Hubble flow and starts to collapse onto the wake. This means that $\dot{h} = 0$,
or equivalently, $x + 2 \psi(x, t) =0$. This yields
\begin{equation}
\label{29}
x(t) = - \frac{6}{5}\left[ \frac{u_i t_{i}^{2}}{t} - u_i t_i \left(\frac{t}{t_i} \right)^{\frac{2}{3}} \right] \;\; .
\end{equation}
Calculating now the thickness $d(t)$ and the surface density $\sigma(t)$ of the wake, we have, respectively:
\begin{eqnarray}
\label{30}
d(t) & = & \frac{12}{5} u_i \left[
\frac{t^{4/3}}{t_i^{1/3}} - \frac{t_i^{4/3}}{t^{1/3}}\right] , \nonumber \\
\sigma(t) & = & \rho(t) d(t) = \frac{2u_i}{5\pi G_0 t^{2}}
\left[  \frac{t^{4/3}}{t_i^{1/3}} - \frac{t_i^{4/3}}{t^{1/3}} \right] ,
\end{eqnarray}
where we have used the average density $\rho(t) =
\frac{1}{6\pi G_0 t^2}$ in the matter-dominated era for a flat  Universe.
Clearly, we see that wakes which are formed at $t_i \sim t_{eq}$ have
largest surface density \cite{vacha2} and (21) reduces to
\begin{equation}
\sigma(t) \approx \frac{2u_i}{5\pi G_0 t}
\left( \frac{t}{t_{i}} \right)^{1/3} .
\end{equation}
Replacing expression (6) for $u_i$ in (22), we finally have:
\begin{equation}
\sigma(t) \approx \frac{8}{5} \frac{\mu}{t}
\left( \frac{t}{t_{i}} \right)^{1/3} \left[ 2 v_s \gamma + \frac{\alpha^2}{v_s \gamma} \right] ,
\end{equation}
where the second term in r.h.s. is the dilaton's contribution to the wake's
surface density and, as we have already seen in section 2, this term is
more than 230 times less than the purely geometrical contribution.

\section{Discussion and Conclusions}

The aim of this work was to make a first step toward the understanding of  the formation
and evolution of cosmic string wakes in the context of a scalar-tensor gravity. For this purpose, in section 3,
we studied the motion of non-relativistic, collisionless particles in the metric of a scalar-tensor string described in section 2.
The linearized geodesic equations found in section 2 were integrated over the unperturbed trajectories which
allowed us to calculate the density fluctuations inside and outside the formed wake. The presence of the dilaton has interesting  physical consequences.
The dilaton qualitatively alters the particles number density (\ref{19}) with respect to what occurs in General Relativity.
In the latter case, the particles number density crossing the element $\delta \rho$ is just twice the initial density, while
in the former case we have an additional term which
expresses the dilaton's contribution to the generation of wakes.
Analysing expression (\ref{21}) for the density fluctuation outside the wake, again the dilaton alters the previous result obtained in General Relativity. Since $x < X_0$ on the side of the coming fluid, we see that $\delta n$ in this region is negative which expresses an underdensity outside the wakes. The evolution of the wake was investigated using the
Zel'dovich approximation. Thickness, surface density and ``turn around" surfaces were computed.

It was shown that the presence of the dilaton produces an effect which is very similar to the effect of the string's wiggles on the formation of the wakes, albeit our model is the one for an ordinary string. Comparing
numerically both models, it seems that the dilaton's contribution can be
neglected in favour of the wiggles' contribution: For instance,
expression (13) for the inside density fluctuation is $10^3$ orders of magnitude less than its analogue in the wiggly string in General Relativity \cite{vacha}.
However, we believe that this should not be seen with
``pessimist eyes'': the (small) upper bound to the value of $\alpha^2$ in the matter era is just a confirmation that if gravity was really scalar-tensorial in early eras it evoluted to General Relativity in the present era \cite{dam}. At early epochs, it is expected  that the scalar's contribution was of the same order of the tensor's one. Once the dilaton still contributes to the wake's surface density and this contribution may lead to interesting cosmological consequences. 

Wakes produced by moving strings can provide an 
explanation for filamentary and sheet-like structures 
observed in the Universe \cite{vac}. The wake produced by 
the string in one Hubble time has the shape of a strip of width $\sim v_s t_i$. With the help of the surface density (23), we can easily compute the wake's linear 
mass density, say $\tilde{\mu}$, 
\begin{equation}
\tilde{\mu} \approx \frac{8\mu}{5}\left(\frac{t}{t_i}\right)^{2/3}\left[2v_s \gamma + \frac{\alpha^2}{\gamma} \right] ,
\end{equation}
where we see that the dilaton's contribution independ on the string velocity. If the string moves slower or if we extrapolate our results to earlier epochs of the Universe when the parameter $1/\omega$ has been much larger, we 
conclude that the second term in r.h.s. of eq. (24) will dominate over the GR term. Therefore, the dilatonic wake would have direct implications on the formation of large-scale structure in the Universe.

\section*{Acknowledgments}
This work was partially supported by the Funda\c{c}\~ao de Apoio \`a Pesquisa do Distrito Federal (FAPDF). SRMM thanks to CNPq for a PhD grant. 
The authors would like to thank the referees for very interesting comments 
and suggestions on the previous version of this manuscript.


\begin{thebibliography}{99}

\bibitem{kibble}Kibble T 1976 {\it J.Phys. A} {\bf 9} 1387,
Vilenkin A and Shellard E P S 1994 Cosmic string and other topological defects {\it Cambridge University Press}.
\bibitem{vil}Silk J and Vilenkin A 1984 {\it Phys. Rev. Lett.}{\bf53} 1700.
\bibitem{vacha2}Vachaspati T 1986
{\it Phys. Rev. Letters} {\bf 57} 1655.
\bibitem{steb}Stebbins A, Veeraraghavan S,
Brandenberger R H, Silk J  and  Turok N 1987
{\it Astrophys. J.} {\bf 1} 322.
\bibitem{deruelle}Deruelle N and  Linet B 1988 {\it Class. Quantum Grav.} {\bf 5} 55.
\bibitem{hiscock}Hiscock W A and Lail B 1988 {\it Phys. Rev. D} {\bf 37} 869.
\bibitem{peri}Perivolaropoulos L, Brandenberger R H and Stebbins A 1990 {\it Phys. Rev. D} {\bf 41} 1764.
\bibitem{vacha}Vachaspati T 1992 {\it Phys. Rev. D} {\bf 45} 3487.
\bibitem{gund}Gundlach C and Ortiz M E 1990 {\it Phys. Rev. D} {\bf 42} 2521, Pimentel L O  and   No\'e Morales A 1990
{\it Rev. Mexicana de F\'{\i}sica} {\bf 36} S199.
\bibitem{br}Barros A and Romero C 1995 {\it J. Math. Phys.}
{\bf 36} 5800.
\bibitem{gregory}Gregory R and  Santos C 1997 {\it Phys. Rev. D} {\bf 56} 1194.
\bibitem{guima}Guimar\~aes M E X 1997 {\it Class. Quantum Grav.} {\bf 14} 435.
\bibitem{sen}Sen A A  1999 {\it Phys. Rev. D.} {\bf 60} 067501.
\bibitem{mexg}Ferreira C N, Guimar\~aes M E X and Helay\"el-Neto, J A  2000 {\it gr-qc/0002054, Nuclear Physics B, to appear}.
\bibitem{bergmann}Bergmann P G 1986 {\it Int. J. Theor. Phys.} {\bf 1} 25.
\bibitem{wagoner}Wagoner R V 1970 {\it Phys. Rev. D} {\bf 1} 3209.
\bibitem{nordtverdt}Nordtverdt Jr K 1970 {\it Astrophys. J.}
{\bf 161} 1059.
\bibitem{dam}Damour Th and Nordtverdt K 1993 {\it Phys. Rev. D} {\bf 48} 3436.
\bibitem{bd}Brans C and Dicke  R H 1961 {\it Phys. Rev.} {\bf 124} 925.
\bibitem{eub}Eubanks T M {\em et al.} 1997 {\it Bull. Am. Phys. Soc.} 
{\bf Abstract \# K 11.05}.
\bibitem{allen}Allen B and Shellard E P S 1990 {\it Phys. Rev Lett.} {\bf 64} 119.
\bibitem{vac}Vachaspati T 1986 {\it Nucl. Phys.}{\bf B277} 593. 
\end{thebibliography}
\end{document}